\documentclass[,final]
  {aipproc}
\layoutstyle{6x9}
\begin{document}

\title 
      {Spectral Properties of M87 Using Two-Component Flow}


\author{Samir Mandal}{
  address={Indian Centre for Space Physics, 43 Chalantika, Garia Station Road, Kolkata 700084, India},
  email={samir@csp.res.in}
}

\begin{abstract}
We fit the observational data for M87 using two-component advective disk model.
We show that the flat spectrum from the nucleus of M87 is due to synchrotron radiation produced by non-thermal 
electrons in the CENBOL. The non-thermal distribution is produced due to acceleration of electrons across the shock 
in a sub-Keplerian flow. 
\end{abstract}
\date{\today}
\maketitle

\section{Introduction}
The accretion flow around a black hole may have two components: 
the optically thick and geometrically thin Keplerian accretion disk 
\cite{SS} on the equatorial plane and an optically thin sub-Keplerian flow \cite{CT,CM} 
on the top of the Keplerian disk. The sub-Keplerian flow can have a shock in presence of angular momentum and 
the post-shock region is hot since flow kinetic energy is converted into thermal energy due to shock compression. 
This hot post-shock region is known as the CENBOL (CENtrifugal pressure supported BOundary Layer) of the black hole.
These shocks still play an important role
in shaping the spectrum. Hence, it is likely  that the {\it standing shocks},
through which all the accreting matter pass before entering into a black hole,
or forming a jet, should be important to energize matter also.
Particularly important is that the post-shock region which is the
repository of hot electrons, can easily inverse Comptonize the soft photons from
a Keplerian disk located in the pre-shock region and reprocessed them to produce hard X-rays. 
A power-law component of the flow is thus formed without taking resort to any hypothetical electron cloud
originally invoked in the literature. Thus, the CENBOL region in between
the horizon and the shock behaves like a boundary layer where
the flow dissipates its gravitational energy. This boundary layer
can oscillate when the cooling is introduced and this
causes quasi-periodic oscillations in X-rays \cite{CM00}.
However, the shocks have another important role.
Given that the accretion flows largely pass through a standing, oscillating or a moving shock,
it is likely that the hot electrons may be {\it accelerated} by them just as
high energy cosmic rays are produced by the transient super-novae shocks \cite{B78a,B78b}.
This acceleration process energies the electrons very efficiently and produce a power-law distribution. 
The accelerated
particles in turn emit synchrotron radiation in presence of the
magnetic field. We calculate the accretion disk spectrum as a function of the shock strength, compression ratio
and flow accretion rate. In the absence of a satisfactory
description of magnetic fields inside the advective disk, we
consider the presence of only stochastic fields and use the
ratio of the field energy to the gravitational energy density to be
a parameter.
The radiation pressure in the hot CENBOL launches a fraction of disk matter 
in the form of a jet which is common in all systems from quasars to nano-quasars. In particular, the M87 jet  
which is supposed to be generated within few tens of Schwarzschild radii \cite{JBL99} of the central black hole 
supports the CENBOL picture. So, the sub-Keplerian matter is an important component for an accretion disk. 
We calculate the spectrum from accretion disk of M87 following the same approach as 
proposed for the galactic black holes \cite{CM} and we fit the observed data by our model. 

In Section 2, we present the model description and the fitting of data. Finally, in Section 3,
we present our concluding remarks.

\section{Fitting of the Observed Data}
We consider a vertically averaged two-component flow around a Schwarzschild black hole.
The black hole geometry is described by a pseudo-Newtonian potential \cite{PW} and the
vertical height of the disk at any radial distance has been calculated by balancing the vertical component of
gravitational force with the gas pressure \cite{C89}. The radial distance is measured in units of
Schwarzschild radius ($r_g$). The parameters of our model are shock location ($x_s$), compression ratio ($R$),
the fraction of electrons having non-thermal distribution of energy ($\xi$) and the accretion rates
of the Keplerian (${\dot m}_d$) and the sub-Keplerian halo (${\dot m}_h$) components.
These accretion rates are measured in units of Eddington rate.
The supply of the sub-Keplerian mater increases the number density of electrons in the CENBOL and also
enhances the synchrotron emission and it self-Comptonization. This will cool the CENBOL region. On the other hand,
supply of Keplerian matter increases the number of soft photons. 
So, the relative accretion rates of the two-components control the spectral properties of the radiation spectrum from the disk.
We assume that the accretion disk of M87 is purely a sub-Keplerian flow and one hardly needs any contribution from 
the Keplerian disk to fit the data.  
\begin{figure}
  \includegraphics[height=.5\textheight]{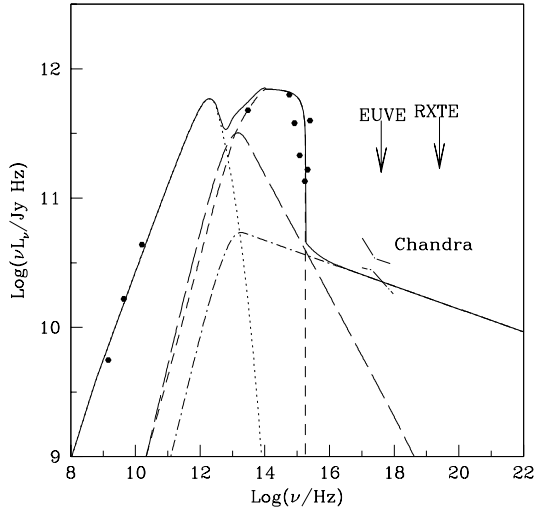}
  \caption{Fitting the spectrum of M87 nucleus using the shock solution of sub-Keplerian flow.
The solid dots are the observational \cite{R04} data points. The dotted line represents
the pre-shock synchrotron contribution while the small dashed line is due to post-shock synchrotron
contribution from non-thermal electrons. The long-dashed and dot-dashed line represents the synchrotron
self-comptonized spectrum due to thermal and non-thermal electrons in the CENBOL.
The solid line represents the total contribution from the accretion disk. }
\end{figure}
 There are several works in the literature which
favor the sub-Keplerian flow model. For example, the low-ionization nuclear emission-line regions
of M87 are produced due to the shock excitation in a dissipative accretion \cite{D97}.
In 1995, Chakrabarti \cite{C95} calculated the velocity field of the ionized disk using spiral shock solution and compare the
shapes of the line profiles expected from various regions of the disk with the HST observation data. The
mass of the central object of M87 was found to be $M = (4 \pm 0.2)\times 10^9 M_\odot$ which is consistent
with the present value. In a sub-Keplerian flow, shocks are very natural outcome and
we have considered that the pre-shock electrons follow a pure thermal distribution while the post-shock flow is a
mixture of thermal and non-thermal electrons because of the acceleration of electrons across the shock.
The slope ($p=\frac{R+2}{R-1})$ of the non-thermal distribution depends on the compression ratio ($R$) 
and it produces a synchrotron spectrum of power-law index [$\alpha=(1-p)/2$].
The minimum energy of the accelerated electrons is derived from the temperature of the pre-shock flow and the
power-law distribution shows a sharp cut-off at maximum energy which derived from the shock location and the
strength of the shock \cite{MC05}. The location of the shock can change depending on the 
specific angular momentum and specific energy of the flow.

In Fig. 1 we presented a preliminary fit of the observational data (solid points) taken from \cite{R04}.
We have chosen the parameters $M = 3.2\times 10^9 M_\odot,\ x_s= 10.0,\ R=2.5,\ 
\xi=0.008,\ {\dot m}_h=0.35$ to fit the data. The dotted line represents the pre-shock synchrotron
contribution while the small dashed line is due to post-shock synchrotron contribution from non-thermal electrons.
The long-dashed and dot-dashed line represents the synchrotron self-comptonized spectrum due to thermal and non-thermal
electrons in the CENBOL.
The disk rate is ${\dot m}_d \le 0.001$ which means that the Keplerian disk is inefficient or absent.
The low energy data fit well by the thermal synchrotron radiation produced by the matter far (pre-shock flow) from the
central object. A shock of compression ratio $R=2.5$ can explain the flat part ($\alpha+1$) of the spectrum and
the sharp cut-off is due to cut-off in the non-thermal distribution itself. 
The synchrotron self-comptonization spectrum in the high energy is consistent with the Chandra data. 
The detection limit of the other instruments in this energy range are mentioned in Fig. 1. Fits with more current data will be presented elsewhere\citep{MC08}.

\section{Concluding Remarks}
We have taken the effects of accretion shocks in calculating the accretion disk spectrum,
in particular, the effect of  the shock in accelerating electrons. Physically,
synchrotron emissions from power-law electrons should also be inverse-Comptonized by the power-law electrons.
We have been able to demonstrate that the CENBOL plays a crucial role in determining spectral features.
The CENBOL is denser and hotter and the electrons obey the power-law distribution. In the post-shock region the magnetic field is stronger and the radiation emitted is more intense.
We show that a fit of the observational data of M87 by a sub-Keplerian flow can be achieved.
The flat spectrum is due to the synchrotron  radiation from non-thermal electrons produced by the 
acceleration across the shock.
In a sub-Keplerian flow the the radial velocity behaves like free fall and the infall time scale 
for M87 in the CENBOL region ($\sim$ tens of $r_g$) is order of few months whereas if we believe that 
the accretion disk of M87 is a Keplerian disk the viscous time scale in the same geometrical region is of 
the order of few hundreds of year. The months scale variability of the core of M87 \cite{P03,H97} in optical/x-ray
wavelength indicates that the radiation comes from a region close to the black hole and this 
region behaves as a sub-Keplerian flow. So, the variability behaviour and the spectral signature indicate a sub-Keplerian
nature of the flow around M87. 
\begin{theacknowledgments}
This work is supported by DST Fast Track Young Scientist Project
\end{theacknowledgments}

\end{document}